\documentclass[aps,prd,twocolumn,nofootinbib,
superscriptaddress,
]{revtex4-1}

\usepackage{hyperref}
\usepackage{url}
\usepackage{setspace}
\usepackage{amsmath,amssymb,amscd,braket}
\usepackage{amsfonts}
\usepackage{color}
\usepackage{graphicx}
\usepackage{xcolor}

\usepackage{lettrine, Romantik}

\newcommand{\bea}{\begin{eqnarray}}
\newcommand{\eea}{\end{eqnarray}}
\newcommand{\be}{\begin{equation}}
\newcommand{\ee}{\end{equation}}
\newcommand{\ba}{\begin{align}}
\newcommand{\ea}{\end{align}}

\begin{document}
\title{
Stable evolution of relativistic hydrodynamics order-by-order in gradients
}

\author{Michal P. Heller} \email{michal.p.heller@ugent.be}
\affiliation{Department of Physics and Astronomy, Ghent University, 9000 Ghent, Belgium}
\affiliation{Institute of Theoretical Physics and Mark Kac Center for Complex
Systems Research, Jagiellonian University, 30-348 Cracow, Poland}

\author{Alexandre Serantes}
\email{alexandre.serantesrubianes@ugent.be}
\affiliation{Department of Physics and Astronomy, Ghent University, 9000 Ghent, Belgium}

\author{Micha\l\ Spali\'nski}
\email{michal.spalinski@ncbj.gov.pl}
\affiliation{National Centre for Nuclear Research, 02-093 Warsaw, Poland}
\affiliation{Physics Department, University of Bia{\l}ystok, 15-245 Bia\l ystok, Poland}

\author{Benjamin Withers}
\email{b.s.withers@soton.ac.uk}
\affiliation{Mathematical Sciences and STAG Research Centre, University of Southampton, Highfield, Southampton SO17 1BJ, UK}

\begin{abstract}
We provide a systematic framework for solving the initial value problem for relativistic hydrodynamics formulated as a gradient expansion.
Secular growth is handled by a suitable covariant resummation scheme, which reorganises the degrees of freedom at each order in the expansion while preserving the sum. 
Our scheme can be applied to any order in the gradient expansion; we provide the explicit formulation at first and second orders.
When working to first order, we find that the BDNK equations of motion emerge as an intermediate step in a calculation performed in the Landau frame.
We show that non-hydrodynamic modes appear only in such intermediate calculations and cancel when evaluating solutions to the required order.
Our procedure does not introduce any other fields or require any additional initial data beyond those appearing in the theory of ideal fluids.
\end{abstract}

\maketitle

\section{Introduction \label{sec.intro}} 
The 21st century has been a golden age for relativistic hydrodynamics. In particular, over the course of the past 25 years, it has established itself as the key framework to model the time evolution of the primordial quark-gluon plasma at ultrarelativistic nuclear collisions at RHIC and LHC~\cite{Heinz:2013th}. The magnitude of the leading dissipative effect associated with the shear viscosity has led to the fundamental insight that the quark-gluon plasma seen in experiments is a strongly-coupled liquid~\cite{Busza:2018rrf}.

Arguably, the majority of the key developments in relativistic hydrodynamics have revolved around dissipative effects, which are clearly crucially important. The challenge that we address in our present work is that of an initial value problem in relativistic hydrodynamics with dissipation. In our view, this is the key foundational problem in this area, as its handling has a bearing on the validity and trustworthiness of predictions of the future behaviour of any relativistic fluid.

The reason why the challenge arises in the first place stems from relativistic hydrodynamics being a classical effective field theory (EFT). Its degrees of freedom are local densities of conserved charges, and its equations of motion are the conservation equations of the currents expressed formally in terms of the densities through a gradient expansion. Including more and more terms in this gradient expansion while preserving covariance adds time derivatives. If these equations of motion are then truncated to a given order, there is no guarantee that the initial value problem will be well posed, and indeed generically it is not \cite{HISCOCK1983466}. The leading order in the gradient expansion defines the theory of ideal fluids that is free from these issues, but it also lacks dissipation.

In the literature, there exist three approaches to deal with this problem. In the M{\"u}ller-Israel-Stewart (MIS) approach, dating back to the 1960s and 1970s~\cite{Muller, Israel1, Israel2}, new transient degrees of freedom are added explicitly, as dissipative corrections to ideal relativistic hydrodynamics are promoted to independent fields obeying relaxation equations. In the much more recent Bemfica-Disconzi-Noronha-Kovtun (BDNK) approach~\cite{Bemfica:2017wps, Bemfica:2019knx, Bemfica:2020zjp, Kovtun:2019hdm}, a field redefinition called a frame transformation and a truncation to first order are combined to generate second-order hyperbolic equations of motion for the charge densities. In this approach, new transient degrees of freedom are introduced implicitly, as the resulting equations of motion involve additional time derivatives relative to the ideal case. Both the MIS and the BDNK approaches are manifestly covariant and both involve transient, exponentially decaying excitations on top of gapless hydrodynamic excitations. Finally, the most recent proposal, known as the density frame approach~\cite{Basar:2024qxd,Bhambure:2024axa,Bhambure:2024gnf}, achieves stability without adding new degrees of freedom at the price of sacrificing relativistic covariance.

Given this diverse set of approaches to dealing with the initial value problem, and the fact that each of them, despite their strengths, has obvious disadvantages, it is natural to wonder if there might be another idea that allows for a fully systematic, order-by-order treatment in the gradient expansion.

In the present work, we go back to the roots and re-examine what it takes to solve the initial value problem in relativistic hydrodynamics as an effective field theory. In other words, we want to solve the initial value problem while working within the gradient expansion up to some desired order of accuracy. 

The naive way of achieving this goal is to first solve the equations of motion for the ideal fluid and then account for the dissipative corrections perturbatively. The power of the formal perturbation parameter, $\epsilon$, can be associated with the order of the gradient term: the ideal terms being ${\cal O}(\epsilon^0)$, the viscous ones ${\cal O}(\epsilon^1)$, etc. The reason why this approach cannot succeed is \emph{secular growth}: a ubiquitous phenomenon in physical systems where the time evolution of initially small perturbative corrections leads to significant growth, thereby invalidating the perturbative expansion. See Ref.~\cite{Craps:2014vaa} for a excellent overview of the topic.

In the present work, we address the problem of secular growth in relativistic hydrodynamics by putting forward a resummation scheme at each order in~$\epsilon$. This method produces a solution to the hydrodynamic conservation equations up to any desired order in~$\epsilon$, and the resulting solution is insensitive to the details of the resummation scheme up to the same order in~$\epsilon$. We first illustrate this idea in Sec.~\ref{sec:diffusion} with the example of diffusion to all orders, and then apply it in Sec .~\ref{sec:rh_1} to a conformal relativistic fluid at first order; in this latter case, we find that the equations of motion of BDNK theory naturally emerge in an intermediate step. To demonstrate the generality of our approach, in Sec.~\ref {sec:rh_2} we work out the equations of motion up to second order, and also outline how the calculation proceeds to any orders in the hydrodynamic gradient expansion.

\section{Diffusion}
\label{sec:diffusion}

The simplest case to consider is charge diffusion. In this case, the conservation equation,
\be
\partial_t\rho + \partial_i j^i = 0, \label{diff:conservation}
\ee
is supplemented with a constitutive relation truncated to first order in the gradient expansion (Fick's law)
\be
j_i = -\epsilon D\partial_i \rho + {\cal O}(\epsilon)^2 \label{diff:constitutive}.
\ee
where $D$ is a constant. Here we are working perturbatively in gradients to first order, with errors of order $\epsilon^2$, where powers of $\epsilon$ count the number of derivatives.

The key characteristic of the EFT is its definition as a perturbative expansion in gradients. Taking this seriously, means a solution to \eqref{diff:conservation} with \eqref{diff:constitutive} should be constructed in a gradient expansion,
\be
\rho(t,\vec{x}) = \rho_0(t,\vec{x}) + \rho_1(t,\vec{x}) \epsilon + {\cal O}(\epsilon)^2, \label{rhoexpansion}
\ee
leading to the following equations of motion at each order in $\epsilon$
\begin{subequations}\label{diffusion}
\begin{align}
\partial_t \rho_0 &= 0, \label{diffusion1}\\
\partial_t \rho_1 &= D \partial^2\rho_0. \label{diffusion2}
\end{align}
\end{subequations}
Direct solution of these equations shows secular growth,
\be
\rho_0 = c_0(\vec{x}),\quad \rho_1 = c_1(\vec{x}) + D \partial^2 c_0(\vec{x}) t,
\ee
where $c_0$, $c_1$ are integration constants.

Our proposed resolution to the issue of secular growth is a field redefinition at the level of the expanded fields, $\rho_0$, $\rho_1$, while leaving $\rho$ invariant to the required order. Near $t=t'$ this is achieved as follows,
\begin{subequations}
\bea
\rho_0 &\to& \rho_0 - \epsilon \alpha \rho_1 \, (t-t'), \label{rhorep0}\\
\rho_1 &\to& \rho_1 + \alpha \rho_1 \, (t-t').\label{rhorep1}
\eea
\end{subequations}
where $\alpha$ is a constant parameter. This has no effect on the solution $\rho$, since
\bea
\rho &=& \rho_0 + \rho _1 \epsilon + {\cal O}(\epsilon)^2\nonumber\\
&\to&(\rho_0 - \epsilon \alpha \rho_1 \, (t-t')) + (\rho_1 + \alpha \rho_1 \, (t-t') ) \epsilon + {\cal{O}}(\epsilon)^2\nonumber\\
&=& \rho.
\eea
We emphasise that this is not a frame transformation since it leaves $\rho$ unaffected.
However, it does change the individual equations of motion for $\rho_0$ and $\rho_1$ due to the time derivatives they contain; in particular after the replacement \eqref{rhorep0}, \eqref{rhorep1} the equations \eqref{diffusion} become 
\begin{subequations}\label{diffusion_resum}
\bea
\partial_t \rho_0 &=& \epsilon \alpha \rho_1, \label{diffusion_resum1}\\
\partial_t \rho_1 &=& D \partial^2\rho_0 - \alpha \rho_1.
\eea
\end{subequations}
To this order the original equation of motion \eqref{diff:conservation} is unaffected by this change, since by \eqref{diffusion_resum}, $\partial_t \rho_0 + \epsilon \partial_t \rho_1 = \epsilon D\partial^2 \rho_0 + \mathcal{O}(\epsilon)^2$ with the $\alpha$ dependence cancelling. Only the equations governing $\rho_0$ and $\rho_1$ are affected, where $\alpha$ is playing the role of an exponential damping rate, and so $\rho_0$ and $\rho_1$ no longer exhibit secular growth. Alternatively, if one were to discretise the time derivatives in \eqref{diffusion_resum}, it is clear that the associated updating scheme involves the transfer of order-1 quantities into order-0 at a rate governed by $\alpha$, so that secular growth is tamed. 
Once \eqref{diffusion_resum} are solved to obtain $\rho_0$ and $\rho _1$, one can reassemble them using \eqref{rhoexpansion} to compute the final result, $\rho$. 

To demonstrate that there is no secular growth we can convert the pair of first-order equations \eqref{diffusion_resum} into a single second-order PDE for $\rho_0$ by eliminating $\rho_1$,
\be
\alpha^{-1}\partial_t^2 \rho_0 +  \partial_t \rho_0 - \epsilon  D \partial^2\rho_0 = 0 \label{telegrapher}
\ee
which is nothing but the telegrapher's equation. It arises both in MIS and BDNK as a shear channel perturbation (viscous diffusion). If one solves \eqref{telegrapher}, then $\rho_1$ is needed in order to obtain $\rho$ through \eqref{rhoexpansion}, and is given simply by the time derivative of $\rho_0$ through \eqref{diffusion_resum1}.

At first glance, if we consider \eqref{diffusion_resum} (or indeed \eqref{telegrapher}) then we are led to the conclusion that 
there are two independent plane-wave solutions (modes) with spatial wave-vector $\vec{k}$, having respective frequencies $\omega_\text{H}$ and $\omega_\text{NH}$. The general solution for $\rho_0$ and $\rho_1$ is then a linear combination of these modes, with coefficients given in terms of initial data for the problem \eqref{diffusion_resum} by
\begin{subequations}
\bea
\rho_0(t) &=& \left(\frac{\alpha + \Delta}{2\Delta} \rho_0(0) + \frac{\alpha \epsilon}{\Delta}\rho_1(0)\right)e^{-i \omega_\text{H} t}\nonumber\\
&& + \left(\frac{-\alpha + \Delta}{2\Delta} \rho_0(0) - \frac{\alpha \epsilon}{\Delta}\rho_1(0)\right)e^{-i \omega_\text{NH} t},\\
\rho_1(t) &=& \left(-\frac{D k^2}{\Delta} \rho_0(0) + \frac{-\alpha +\Delta}{2\Delta}\rho_1(0)\right)e^{-i \omega_\text{H} t}\nonumber\\
&& + \left(\frac{Dk^2}{\Delta} \rho_0(0) + \frac{(\alpha +\Delta)}{2\Delta}\rho_1(0)\right)e^{-i \omega_\text{NH} t},\;\;\;\quad
\eea
\end{subequations}
where $\omega_\text{H} = -\frac{i}{2}(\alpha-\Delta)$,  $\omega_\text{NH} = -\frac{i}{2}(\alpha+\Delta)$ with~$\Delta = \sqrt{\alpha(\alpha - 4 D k^2 \epsilon)}$.
The $\omega_\text{H}$ behaviour is gapless, hydrodynamic, since $\omega \to 0$ as $k\to 0$, while the $\omega_\text{NH}$ behaviour is gapped, hence non-hydrodynamic (transient). 
A key insight that follows is that while $\omega_\text{NH}$ appears in the solution to $\rho_0$ and $\rho_1$, it does not contribute to $\rho$ at the required order in $\epsilon$ since it cancels out in the final sum~\eqref{rhoexpansion},
\be \label{rhocancel}
\begin{aligned}
\rho(t) =& \left(\rho_0(0) + \epsilon \rho_1(0) + {\cal O}(\epsilon)^2\right)e^{-i \omega_\text{H} t}\\
&+ \left({\cal O}(\epsilon)^2\right)e^{-i \omega_\text{NH} t}.
\end{aligned}
\ee
Therefore, to first order in $\epsilon$, the non-hydrodynamic mode does not appear in the final solution, even though it plays a key role in the causal and stable construction of $\rho_0$ and $\rho_1$ via \eqref{telegrapher}. 
The parameter $\alpha$, which in our scheme is present to perform secular resummation, in the context of \eqref{telegrapher} plays the role of a regulator which can be chosen so that the evolution is causal. 
In addition, the initial data at each order in the $\epsilon$ expansion, $\rho_i(0)$, combine to give the physical initial data, i.e. $\rho(0)$. It clearly does not matter how one chooses to distribute $\rho(0)$ between the $\rho_i(0)$ for the purposes of evolution. This emphasises that despite the increased differential order, no additional physical initial data are required.

This construction naturally extends to higher orders. The constitutive relation gets extended to include new transport coefficients, $\beta_{2n}$,
\be
j_i = \sum_{n=1}^\infty \beta_{2n}\epsilon^{2n-1}(-\partial^{2})^{n-1} \partial_i \rho 
\ee
where $\beta_2 = -D$~\cite{Heller:2020uuy}, along with an extension of \eqref{rhoexpansion},
\be
\rho = \sum_{n=0}^\infty \epsilon^n \rho_n.
\ee
The resulting equations of motion can be modified to absorb secular growth, just as going from \eqref{diffusion1}, \eqref{diffusion2} to \eqref{diffusion_resum}. A natural minimal choice of such adjustments introduces a secular resummation parameter $\alpha_n$ at each order, leading to the following equations,
\begin{subequations}
\begin{align}
\partial_t \rho_0 &= \epsilon \alpha_1 \rho_1\\
\partial_t \rho_1 &= -\beta_2 \partial^2 \rho_0  - \alpha_1 \rho_1 + \epsilon \alpha_2 \rho_2\\
\partial_t \rho_2 &= -\beta_2 \partial^2 \rho_1  - \alpha_2 \rho_2 + \epsilon \alpha_3 \rho_3\\
& \vdots \nonumber\\
\partial_t \rho_m &= \sum_{n=1}^{\lfloor\frac{1+m}{2}\rfloor} \beta_{2n}(-\partial^2)^{n} \rho_{1+m-2n}- \alpha_{m} \rho_m + \epsilon \alpha_{m+1} \rho_{m+1}, \label{dtrhom}
\end{align}
\end{subequations}
The spectrum of modes at spatial wave-vector $\vec{k}$ is thus governed by a bidiagonal matrix independent of $k$, plus a lower-triangular Toeplitz matrix dependent on $k$.
At $k=0$, the spectrum is given by the bidiagonal matrix, whose eigenvalues are the diagonals, 
\be
-i\omega_n = -\alpha_n, \qquad n = 0, 1, \ldots
\ee
with $\alpha_0 = 0$, corresponding to one gapless excitation (the hydrodynamic mode) and infinitely many gapped modes. The eigenvector corresponding to the $n$th eigenvalue, $\vec{\rho}^{(n)}$, is then be obtained by solving \eqref{dtrhom} with the eigenvalue substituted in, i.e.
\be
\rho^{(n)}_{m+1} = \frac{\alpha_m-\alpha_n}{\epsilon \alpha_{m+1}} \rho^{(n)}_m,
\ee
with solution
\be
\rho_m^{(n)} = \left(\prod_{j=0}^{m-1}\frac{\alpha_j-\alpha_n}{\epsilon\alpha_{j+1}}\right)\rho_0^{(n)}, \qquad m\geq 1.
\ee
i.e. the $n$-th eigenfunction has the following $\epsilon$-expansion
\bea
\rho^{(n)} &=& \sum_{m=0}\rho_m^{(n)}\epsilon^m\nonumber\\
&=& \rho_0^{(n)}\left(1 + \sum_{m=1}\prod_{j=0}^{m-1}\frac{\alpha_j-\alpha_n}{\alpha_{j+1}}\right)
\eea
First, we establish that $\rho^{(n>0)} = 0$, generalising the second line of \eqref{rhocancel}. This follows directly from the identity\footnote{This can be proven as follows. Let $P_m = \prod_{j=0}^{m-1} \frac{\alpha_j-\alpha_n}{\alpha_{j+1}}$ and $P_0 = 1$. Then by induction one can prove that $P_k = - \frac{\alpha_n}{\alpha_k}\sum_{s = 0}^{k-1} P_s$. Hence $P_n = -\sum_{s = 0}^{n-1} P_s$ from which \eqref{sumprodid} follows.
}
\be\label{sumprodid}
\sum_{m=1}\prod_{j=0}^{m-1}\frac{\alpha_j-\alpha_n}{\alpha_{j+1}} = -1, \qquad n > 0.
\ee
Next, we can show that $\rho^{(0)}$ is simply the appropriate sum of initial data, since
\bea
\rho^{(0)} &=& \sum_{m=0} \rho_m^{(0)}\epsilon^m\nonumber\\
&=& \sum_{n=0} \sum_{m=0} \rho_m^{(n)}\epsilon^m e^{-i\omega_n t}\bigg|_{t=0}\nonumber\\
&=& \sum_{m=0} \rho_m(0)\epsilon^m
\eea
where in the second line we used the property that $\rho^{(n>0)} = 0$, and in the third line we introduced $\rho_m(0)$ as the initial data at $t=0$ for $\rho_m$. This generalises the first line of \eqref{rhocancel} to all orders.

\section{Hydrodynamics to first order}
\label{sec:rh_1}

In this section, we consider the hydrodynamic description of a neutral relativistic fluid, for which the hydrodynamic variables are the local temperature $T(x)$ and the local fluid velocity $U^\mu(x)$. The stress-energy tensor of the fluid is parametrised in terms of these variables through the gradient expansion, 
here shown to first order in gradients in the Landau frame:
\begin{align}
\label{stress-energy}
T_{\mu\nu} &= \mathcal{E} U_\mu U_\nu + \mathcal{P} \Delta_{\mu\nu}\nonumber\\
&- \epsilon\, \eta\, \sigma_{\mu\nu} - \epsilon\,\zeta\, \Delta_{\mu\nu} \partial\cdot U + {\cal O}(\epsilon)^2, 
\end{align}
where as before, $\epsilon$ is a formal parameter to count derivatives. We have introduced the spatial projector $\Delta_{\mu\nu} = \eta_{\mu\nu} + U_\mu U_\nu$, and the shear tensor is given by
\be
\sigma_{\mu\nu} = \Delta_\mu^{~a}\Delta_\nu^{~b}\left(\partial_a U_b + \partial_b U_a - \frac{2}{d-1}\eta_{ab} \partial\cdot U\right).
\ee
The equation of motion of relativistic hydrodynamics are the conservation of \eqref{stress-energy}, $\partial_\mu T^{\mu\nu}=0$. 

For the sake of presentational simplicity, we specialise to the conformal case where $\zeta = 0$, $\mathcal{P} = \mathcal{E}/(d-1)$, and we define the local temperature $T$ in terms of the energy density $\mathcal{E}$ through the equation of state, $\mathcal{E} = c_\mathcal{E} T^{d}$. For such a conformal fluid, the shear viscosity is given by $\eta(T) = \frac{\eta}{s} \frac{d}{d-1}c_{\mathcal{E}}T^{d-1}$, where the specific shear viscosity $\eta/s$ is a dimensionless number.  

The first step is to expand the fields in the gradient expansion, using the formal parameter $\epsilon$, 
\be\label{ansatz}
T(x) = \sum_{n=0}^\infty T_n(x) \epsilon^n, \quad U^\mu(x) = \sum_{n=0}^\infty U^\mu_n(x) \epsilon^n. 
\ee
Note that the normalisation of the fluid velocity, $U\cdot U = -1$, imposes an infinite set of nontrivial algebraic relations on the $U_n^\mu$ fields. At the three lowest orders, 
\bea
U_0\cdot U_0 = -1, \quad U_0\cdot U_1 = 0, \quad 2 U_0\cdot U_2 + U_1\cdot U_1 = 0.\nonumber\\\label{orthoU}
\eea

The ansatz \eqref{ansatz} leads to the naive perturbation theory equations of motion to first order in gradients, 
\begin{subequations}\label{hydro1unmodified}
\bea
\partial_\mu T_0^{\mu\nu}  &=& 0,\\
\partial_\mu T_1^{\mu\nu}  &=& 0,
\eea
\end{subequations}
which govern the variables $T_0, U_0^\mu$ and $T_1, U_1^\mu$. For future reference, 
\be
T^0_{\mu\nu} = \mathcal{E}_0\left(U^0_\mu U^0_\nu + \frac{1}{d-1}\Delta^0_{\mu\nu} \right).
\ee
The next step is the secular resummation step, in which we make the following replacements at the level of the equations of motion,\footnote{The $U^\mu$ rule here can be induced from a rule on the underlying spatial velocity, $\vec{v}$.}
\begin{subequations}\label{resummation_1}
\bea
T_0 &\to & T_0 + \epsilon\alpha T_1 U_0\cdot (x-x'),\\
T_1 &\to & T_1 - \alpha T_1 U_0\cdot (x-x'),\\
U_0^\mu &\to& U_0^\mu + \epsilon \beta U_1^\mu\,U_0\cdot (x-x'),\\
U_1^\mu &\to& U_1^\mu - \beta U_1^\mu\,U_0\cdot (x-x') + {\cal O}(\epsilon),
\eea
\end{subequations}
at each location $x'$, and then take $x'\to x$ at the end. This has the interpretation of transferring secularly growing pieces at order $\epsilon^1$ into order $\epsilon^0$ terms, just as in the diffusion example, here controlled by the parameters $\alpha$ and $\beta$. These rules preserve the normalisation conditions \eqref{orthoU}. The overall effect of these replacements is to adjust material derivative terms ($U_0^\mu\partial_\mu$) in \eqref{hydro1unmodified}, leading to the following set of coupled first-order equations,
\begin{subequations}\label{resumJ}
\bea
\partial_\mu T_0^{\mu\nu}  &=& \epsilon\, J^\nu, \label{resumJ1}\\
\partial_\mu T_1^{\mu\nu}  &=& -J^\nu + {\cal O}(\epsilon), \label{resumJ2}
\eea
\end{subequations}
where
\bea
J^\nu &=& -\alpha T_1^{\mu\nu}U^0_\mu + (\beta-\alpha)d\, T_0^{\mu\nu}U^1_\mu. \label{J1def}
\eea
This allows for energy fluxes (controlled by $\alpha$) and momentum fluxes (controlled by $\beta-\alpha$) between order $\epsilon^1$ and $\epsilon^0$ while maintaining overall conservation, $\partial_\mu \left(T_0^{\mu\nu} + \epsilon T_1^{\mu\nu}\right) = {\cal O}(\epsilon)^2$. We note that $\alpha, \beta$ have dimensions of inverse length, and thus for a conformal fluid are taken to be proportional to $T_0$.

In fact, \eqref{resumJ} is a first-order formulation of BDNK in the variables $U_0^{\mu}$ and $T_0$. To see this, note that the first-order variables $U_1^{\mu}$ and $T_1$ can be obtained as derivatives of the zeroth-order ones by projecting \eqref{resumJ1} with $U^0_\nu$ and $\Delta^0_{\rho\nu}$, resulting in
\begin{subequations}\label{T1U1asderiv}
\bea
T_1 &=& -\frac{T_0^{1-d}}{\epsilon d c_\mathcal{E}\alpha} U^0_\nu \partial_\mu T_0^{\mu\nu}, \label{T1asderiv}\\
U^1_\rho &=& \frac{(d-1)T_0^{-d}}{\epsilon d c_\mathcal{E}\beta}\Delta^0_{\rho\nu} \partial_\mu T_0^{\mu\nu}. \label{U1asderiv}
\eea
\end{subequations}
Then the full stress-energy tensor at first order, $T_{\mu\nu} = T^0_{\mu\nu} + T^1_{\mu\nu}\epsilon$, takes the form
\be
T_{\mu\nu} = T^0_{\mu\nu}  - \eta_0 \epsilon\sigma^0_{\mu\nu} + \frac{1}{\alpha} \Pi^1_{\mu\nu} + \frac{1}{\beta} \Pi^2_{\mu\nu},
\ee
where
\begin{subequations}
\bea
\Pi^1_{\mu\nu} &=& \frac{d}{d-1} \left(\partial\cdot U_0 + \frac{d-1}{d} \frac{U_0\cdot\partial \mathcal{E}_0}{\mathcal{E}_0}\right)T^0_{\mu\nu},\\
\Pi^2_{\mu\nu} &=& \frac{d\mathcal{E}_0}{d-1} U^0_{\mu}   \left(U_0\cdot \partial U^0_{\nu} + \frac{1}{d}\frac{(\Delta_0)_{\nu}^{\sigma}\partial_\sigma \mathcal{E}_0}{\mathcal{E}_0}\right) + (\mu \leftrightarrow \nu),\nonumber\\
&&
\eea
\end{subequations}
and where $\mathcal{E}_0, \Delta_0, \sigma_0^{\mu\nu}$, and $\eta_0$ are these quantities evaluated the zeroth order fields.
This is the BDNK stress-energy tensor in the variables $U_0^\mu$ and $T_0$ with $\alpha$ and $\beta$ playing the role of frame parameters. As has been established in BDNK literature, the BDNK equations of motion are well-posed and causal \cite{Bemfica:2017wps,Bemfica:2019knx,Bemfica:2020zjp}. 

The spectrum for $U_0$ and $T_0$ perturbations, as governed by these BDNK equations, contains transient modes. In the shear channel there is a hydrodynamic mode and a non-hydrodynamic mode with gap $\beta$ (i.e. $\omega_\text{NH}(k=0) = -i\beta$),  while in the sound channel there are two hydrodynamic modes and two non-hydrodynamic modes with gaps $\alpha$ and $\beta$.

Note however that here the BDNK equations are simply an intermediate step. The final answer requires taking the BDNK solution and computing $U_1^\mu$ and $T_1$ through \eqref{T1asderiv} and \eqref{U1asderiv}, in order to assemble the full solution for the fields $U^\mu$ and $T$ to the required order. This step, which is not taken in the previous literature, is essential for the cancellation of the gapped degrees of freedom, as is the case in the diffusion example of the last section.\footnote{In fact, the linearisation of this system in the shear channel is the diffusion example.} Additionally, we note that the final result resulting from this procedure is in the Landau frame. 

While going to second order form makes a direct connection to previous literature (BDNK), we note that it is more convenient to remain in the first order form \eqref{resumJ}, where the above considerations are manifest. This will also be convenient for the purposes of going to higher orders in the expansion, which we present in the next section.

\section{Hydrodynamics at higher orders}
\label{sec:rh_2}

At second order in gradients, we can consider the following natural generalization of the resummation rules~\eqref{resummation_1}, 
\begin{subequations}\label{resummation_2}
\bea
T_0 &\to & T_0 + \epsilon\alpha_1 T_1 U_0\cdot(x{-}x'),\\
T_1 &\to & T_1 - \left(\alpha_1 T_1 - \epsilon\alpha_2 T_2 \right)U_0\cdot(x{-}x'),\\
T_2 &\to & T_2 - \alpha_2 T_2 U_0\cdot(x{-}x'),\\
U_0^\mu &\to& U_0^\mu + \epsilon \beta_1 U_1^\mu\,U_0\cdot (x{-}x'),\\
U_1^\mu &\to& U_1^\mu + \left(-\beta_1 U_1^\mu + \epsilon \beta_2 U_2^\mu  + \epsilon \mathcal{X} U_0^\mu\right)U_0\cdot (x{-}x'),\quad\quad\\
U_2^\mu &\to& U_1^\mu + \left(-\beta_2 U_2^\mu - \mathcal{X} U_0^\mu \right)U_0\cdot (x{-}x') + {\cal O}(\epsilon),
\eea
\end{subequations}
where $\alpha_2$ and $\beta_2$ are two new resummation parameters, and $\alpha_1 = \alpha$, $\beta_1 = \beta$ as used in Sec.~\ref{sec:rh_1}. Here $\mathcal{X} = \beta_1 U_1 \cdot U_1 + \beta_2 U_0 \cdot U_2$ which is required to preserve the normalisation relations \eqref{orthoU}.

As at first order, the net effect of these resummation rules is a modification of the equations of motion $\partial_\mu T_0^{\mu\nu} = \partial_\mu T_1^{\mu\nu} = \partial_\mu T_2^{\mu\nu} = 0$ that ultimately amounts to a change of their right-hand sides through the appearance of two currents $J_1^\mu$ and $J_2^\mu$, 
\begin{subequations}\label{eoms_2}
\bea
\partial_\mu T_0^{\mu\nu}  &=& \epsilon\, J_1^\nu, \\
\partial_\mu T_1^{\mu\nu}  &=& -J_1^\nu + \epsilon J_2^\nu,\\
\partial_\mu T_2^{\mu\nu}  &=& -J_2^\nu + {\cal O}(\epsilon),
\eea
\end{subequations}
where $J_1^\nu = J^\nu$ is as given in \eqref{J1def}, and $J_2^\nu$ is given by
\begin{subequations}
\begin{align}    
&J_2^\mu = \mathcal{A}_0 U_0^\mu + \mathcal{A}_1 U_1^\mu + \mathcal{A}_2 U_2^\mu + \mathcal{B}^\mu, \\
&\mathcal{A}_0 = \mathcal{E}_0 \left[d(d-1)\alpha_1 \left(\frac{T_1}{T_0}\right)^2 + d \alpha_2 \frac{T_2}{T_0} \right.\nonumber\\
&\left.\quad\quad+\frac{d (4\beta_1 - \beta_2)}{2(d-1)}U_1\cdot U_1\right],
\\
&\mathcal{A}_1 = \frac{d^2 \mathcal{E}_0 (\alpha_1 +\beta_1)}{d-1} \frac{T_1}{T_0} + \frac{(d+1)\beta_1}{d-1}\eta_0 \partial \cdot U_0,
\\
&\mathcal{A}_2 = \frac{d \mathcal{E}_0 \beta_2}{d-1},
\\
&\mathcal{B}^\mu = -\frac{(d+1)\beta_1}{d-1}\eta_0 U_1^\nu \partial_\nu U_0^\mu. 
\end{align}
\end{subequations}
The linear spectrum for $T_0$ and $U_0^\mu$ perturbations in the shear channel contains one hydrodynamic mode, and two gapped modes with gaps $\beta_1, \beta_2$. In the sound channel this contains two hydrodynamic modes and four gapped modes with gaps $\alpha_1, \alpha_2, \beta_1, \beta_2$. Thus while these parameters should be chosen appropriately for stability and causality of intermediate calculations,\footnote{A preliminary investigation shows that this is indeed possible.} these extra gapped degrees of freedom will not appear in the physical spectrum (i.e. perturbations of $T,U^\mu$).

At higher orders, the resummation rules \eqref{resummation_2} can be straightforwardly generalised, keeping only terms linear in $U_0\cdot(x-x')$ as a scheme choice. As in the all-orders diffusion example, there are new $\alpha_n, \beta_n$ parameters at each order, and additional terms to ensure the preservation of $U_\mu U^\mu = -1$. Then, there is an extension of the first-order time-evolution formalism \eqref{eoms_2} to include an additional current $J_n^\mu$ at each order in $\epsilon$. These $J_n^\mu$ are completely determined by the resummation scheme. Thus, structurally, there is no obstruction to applying our scheme to arbitrary order in the hydrodynamic gradient expansion. In practice, well-posedness would need to be further investigated.

\section{Outlook}

We have formulated a systematic procedure to evolve initial data based on the hydrodynamic gradient expansion. This procedure is free of the secular divergences that have disqualified such approaches in the past. It does not introduce any degrees of freedom nor require any initial data beyond those present in the ideal fluid theory and is not limited to first order in gradients. 

At first order in gradients, our approach leads to the BDNK evolution equations as an intermediate step. The well-posedness of the BDNK equations of motion naturally implies that our approach is correspondingly well-posed. However, the final result of our approach for the hydrodynamic fields $T, U^\mu$ is not the BDNK result, since the additional step involved in computing $T_1$ and $U_1^\mu$ is crucial for obtaining the correct physical answer. In particular, our solution is insensitive to the transient modes that appear in BDNK to the desired order in the gradient expansion, while solutions of BDNK are not. 

Moreover, while it naively seems like there is extra initial data required for evolution compared to ideal hydrodynamics, this is not the case in our approach. This is because the additional fields $U_n^\mu, T_n$ are simply coefficients in an $\epsilon$ expansion of the physical fields, $U^\mu, T$, and thus there is a freedom, without consequence, to choose how the physical initial data is distributed among the different $\epsilon$ orders.

Our approach is of immediate physical relevance to applications of relativistic hydrodynamics in real-world domains like QGP modelling and neutron star mergers. The formalism introduced here can be implemented with a straightforward modification of BDNK codes, since a step in our method can be performed using BDNK equations of motion (obtaining $T_0, U_0^\mu$), and then the final physical solution requires adding $T_1, U_1^\mu$ which is obtained simply by taking a derivative of the BDNK result via~\eqref{T1U1asderiv}. 
Applications to physical phenomena might require incorporating conformal symmetry breaking or coupling to dynamical gravity, which appear to be straightforward extensions. A more challenging research question is the incorporation of fluctuations following a similar effective field theory philosophy.

Going further, our approach generalises to higher orders in gradients, and is naturally formulated in first-order form, as is convenient for implementation of the initial value problem. The first-order form is given by a simple expression involving a new current $J^\mu$ responsible for the transfer of secular growth from higher to lower orders. We have not yet investigated the well-posedness of working to higher orders in gradients.

At the fundamental level, an approach to solving the initial value problem in the context of relativistic hydrodynamics should be causal. Indeed, whilst we did not demand causality when we introduced our secular resummation scheme, the resummation parameters play the role of regulators whose values should be chosen so that the result is causal. For example, the intermediate equation for diffusion is the telegrapher's equation with parameter $\alpha$ which must obey appropriate causality bounds. In first-order hydrodynamics, the intermediate equation is BDNK, which again is causal for an appropriate choice of parameters $\alpha, \beta$. A complete set of results for imposing causality and stability at the linear level can be found in Refs.~\cite{Heller:2022ejw, Heller:2023jtd}. Note that ultimately the dependence on  these parameters formally cancels order by order in the gradient expansion. 

Our framework for treating the initial value problem in hydrodynamics should be generalisable to a broad class of effective field theories. As a first step, it would be instructive to revisit the scalar model of~\cite{Reall:2021ebq,Figueras:2025ofh} through the lens of our approach. Further, one may be interested in addressing higher-derivative theories of gravity, where the initial value problem is particularly relevant and where related formulations have been motivated by MIS~\cite{Cayuso:2023xbc} and BDNK~\cite{Figueras:2024bba}. Extending our perspective to such settings could provide valuable insights for the ongoing tests of general relativity using gravitational-wave observatories.

\begin{acknowledgments}
It is a pleasure to thank Luis Lehner for comments on an initial draft.
This project has received funding from the European Research Council (ERC) under the European Union’s Horizon 2020 research and innovation programme (grant number: 101089093 / project acronym: High-TheQ). Views and opinions expressed are however those of the authors only and do not necessarily reflect those of the European Union or the European Research Council. Neither the European Union nor the granting authority can be held responsible for them. This work was partially supported  by the Priority Research Area Digiworld under the program Excellence Initiative  - Research University at the Jagiellonian University in Krakow. MS was supported by the National Science Centre, Poland, under Grant No. 2021/41/B/ST2/02909. BW is supported by a Royal Society University Research Fellowship and in part by the STFC consolidated grant ‘New Frontiers In Particle Physics, Cosmology And Gravity’.
\end{acknowledgments}

\bibliography{refs} 

\end{document}